\documentclass[showpacs,aps,graphicx,twocolumn]{revtex4}%and
\usepackage{graphicx}
\begin{document}
%\begin{CJK*}{GBK}{song}

\title{Multiparty Quantum Remote Secret Conference\footnote{Supported by the National Natural
Science Foundation of China under Grant Nos. 10604008
and 10435020, and Beijing Education Committee under Grant
No. XK100270454.}}

\author{Xi-Han Li$^{1,2}$, Chun-Yan Li$^{1,2}$,
 Fu-Guo Deng$^{1,2,3}$\footnote{To whom correspondence should be addressed. Email address:
fgdeng@bnu.edu.cn}, Ping Zhou$^{1,2}$, Yu-Jie Liang$^{1,2}$
 and Hong-Yu Zhou$^{1,2,3}$ }
\address{$^1$ The Key Laboratory of Beam Technology and Material
Modification of Ministry of Education, Beijing Normal University,
Beijing 100875,  People's Republic of China\\
$^2$ Institute of Low Energy Nuclear Physics, and Department of
Material Science and Engineering, Beijing Normal University, Beijing
100875,  People's Republic of China\\
$^3$ Beijing Radiation Center, Beijing 100875,  People's Republic of
China}
\date{\today }

\begin{abstract}
We present two schemes for multiparty quantum remote secret
conference in which each legitimate conferee can read out securely
the secret message announced by another one, but a vicious
eavesdropper can get nothing about it. The first one is based on the
same key shared efficiently and securely by all the parties with
Greenberger-Horne-Zeilinger (GHZ) states, and each conferee sends
his secret message to the others with one-time pad crypto-system.
The other one is based on quantum encryption with a quantum key, a
sequence of GHZ states shared among all the conferees and used
repeatedly after confirming their security. Both these schemes are
optimal as their intrinsic efficiency for qubits approaches the
maximal value.

\end{abstract}
\pacs{03.67.Hk, 03.65.Ud} \maketitle

Quantum communication supplies some novel ways for transmitting
message securely. For example, quantum key distribution (QKD)
\cite{Gisin,BB84,BBM92,longqkd,ABC,licy}, the original application
of quantum mechanics, can be used to create a private key between
two authorized users, Alice and Bob. The noncloning theorem forbids
a vicious eavesdropper, say Eve to copy an unknown quantum state
without disturbing it. By analysing the error rate of samples chosen
randomly, Alice and Bob can determine whether there is an
eavesdropper in the quantum line \cite{Gisin}. With quantum secret
sharing (QSS) \cite{HBB99,KKI,Guo,longQSS,BidQSS,Zhang,yanpra}, a
boss can generate a private key with his agents, i.e., $K_A=K_B
\oplus K_C \oplus K_D \oplus \cdots$. Here $K_A$ is the key of the
boss Alice, $K_i$ ($i=B,C,D,\cdots$) are the keys of Alice's agents.
In this way, Alice can send her secret message to her agents who can
read out it if and only if they cooperate, otherwise none can obtain
a useful information about the message. Also, QSS provides a secure
way for sharing an unknown state
\cite{Peng,dengQSTS,dengcontrolled,Lixhjpb}.

Recently, a new branch of quantum communication, quantum secure
direct communication (QSDC) was proposed and has been actively
pursued
\cite{two-step,QOTP,Wangc,Wangcoc,zhangzj,Lixhnetwork,yan,zhangsPRA,cai,song,qcryption2,wangj}.
QSDC makes a party communicate another one directly and securely,
different from QKD. In detail, the sender sends his secret message
directly to the receiver without creating a private key and then
encrypting the message with it. As pointed out in Ref.
\cite{dengnetwork}, a secure quantum direct communication protocol
requires the users to transmit the quantum states in a date block.
Moreover, the two legitimate users can detect the eavesdropper
before they encode the secret message on the quantum states, and can
read out the message directly without exchanging an additional
classical bit for each qubit except for those used for
checking eavesdropping. In this way, the two-step QSDC protocol
\cite{two-step} and the quantum one-time pad QSDC protocol
\cite{QOTP} proposed by Deng \emph{et al.} satisfy all the requirements. So do the protocol with
quantum superdense coding \cite{Wangc} and that with multi-particle
Green-Horne-Zeilinger (GHZ) state \cite{Wangcoc}.

Another class of quantum communication protocols used to transmit
secret message is called deterministic secure quantum communication
(DSQC) \cite{lixhjp}, such as the schemes \cite{yan,zhangzj} with
quantum teleportation and entanglement swapping, and those
\cite{zhangsPRA,wangj}  based on secret transmitting order of
particles. Although the secret message can be read out only after
the two authorized users exchange an additional classical bit for
each qubit, none of the users needs to transmit the qubits that
carry the secret message, which maybe make those protocols more
secure than others in a noise channel and more convenient for
quantum error correction \cite{lixhjp}. In particular, we introduced
a DSQC protocol with only single-photon measurements and a feasible
quantum signal, nonmaximally entangled states \cite{lixhjp}.

More recently, two novel concepts for direct communication are
proposed. One is quantum secret report in which many agents report
directly their secret messages to a boss in one-way direction
\cite{dengreport}. The other is quantum broadcast communication
\cite{wangbroadcast} with which one can broadcast his secret message
to many legitimate receivers. In this Letter, we will present two
schemes for multiparty quantum remote secret conference (MQRSC) in
which any legitimate conferee can send securely his secret message
to the other legitimate parties who participate in a remote secret
conference, but a vicious eavesdropper can get nothing about the
messages. The first one is based on the same classical key shared
efficiently and privately by all the parties with
Greenberger-Horne-Zeilinger (GHZ) states. The other one is based on
quantum encryption with a quantum key, a sequence of GHZ states
shared among all the parties and used repeatedly after confirming
their security.

Now, let us describe the principle of our MQRSC protocol based on
the same private key shared by all the parties of the secret
conference. Suppose there are three remote conferees, say Alice, Bob
and Charlie. They first share the same key, say $K_A=K_B=K_C$
securely and then use them to encrypt their secret messages with
classical one time-pad crypto-system. Here $K_A$, $K_B$ and $K_C$
are the classical keys obtained by Alice, Bob and Charlie,
respectively. In detail, they can use some three-particle GHZ states
to create their keys efficiently. The GHZ-state is
\begin{eqnarray}
\vert \psi\rangle_{ABC}&=&\frac{1}{\sqrt{2}}(\vert 000\rangle + \vert 111\rangle)_{ABC}, \nonumber\\
&=& \frac{1}{\sqrt{2}}[(\vert +x\rangle_A\vert +x\rangle_B + \vert
-x\rangle_A\vert -x\rangle_B)\vert +x\rangle_C \nonumber\\
&& + (\vert +x\rangle_A\vert -x\rangle_B + \vert -x\rangle_A\vert
+x\rangle_B)\vert -x\rangle_C. \nonumber\\
\nonumber\\\label{GHZcorrelation}
\end{eqnarray}
Here $\vert 0\rangle=\vert +z\rangle$ and $\vert 1\rangle=\vert
-z\rangle$ are the eigenvectors of the measuring basis (MB)
$\sigma_z$, and $\vert +x\rangle=\frac{1}{\sqrt{2}}(\vert 0\rangle +
\vert 1\rangle)$ and $\vert -x\rangle=\frac{1}{\sqrt{2}}(\vert
0\rangle - \vert 1\rangle)$ are those of the MB $\sigma_x$. That is,
a remote conferee, say Alice prepares a three-particle quantum
system in the GHZ state $\vert \psi\rangle_{ABC}$, and then sends
the particle $B$ to Bob and the particle $C$ to Charlie. All the
three conferees choose the MBs $\sigma_z$ and $\sigma_x$ with the
probabilities $1-p$ and $p$ to measure their particles. After
comparing in public their MBs, the three conferees keep their
outcomes obtained when they all choose the MB $\sigma_z$ or the MB
$\sigma_x$. The probabilities that they obtain a correlated result
with the MBs $\sigma_z$ and $\sigma_x$ are $(1-p)^3$ and $p^3$,
respectively.

Alice, Bob and Charlie choose all the outcomes obtained with the MB
$\sigma_x$ and a subset of those with the MB $\sigma_z$ as the
samples to analyse the security of the transmission, similar to
Bennett-Brassard-Mermin 1992 (BBM92) QKD protocol \cite{BBM92} and
the modified Bennett-Brassard 1984 (BB84) QKD protocol \cite{ABC}.
If the transmission is secure, the three conferees can get the same
private key by distilling the remaining outcomes obtained with the
MB $\sigma_z$. Assume that the ratio of the number of sample
particles to that of particles transmitted is $r$. As the symmetry,
$p^3=r/2$ (half of the samples are the outcomes obtained by the
conferees with the MB $\sigma_x$). Then the probability that the
three remote conferees obtain their raw key, $p_{rk}$ without
eavesdropping is
\begin{eqnarray}
p_{rk}=(1-\sqrt[3]{\frac{r}{2}})^3 p_t p_d,
\end{eqnarray}
where $p_t$ and $p_d$ are the  probabilities of the transmission and the detectors, respectively.

In essence, this MQRSC protocol is just a special QKD protocol. That
is, all the conferees generate first the same private key with GHZ
states efficiently and then use it to encrypt and decrypt their
messages. Different from QSS
\cite{HBB99,KKI,Guo,Zhang,yanpra,longQSS,BidQSS}, all the authorized
conferees are honest. They can take the measurement with the MB
$\sigma_z$ on the GHZ states to obtain the same outcomes, and use
all the outcomes obtained with the MB $\sigma_x$ as the samples for
eavesdropping check.

\begin{figure}[!h]%[tpb]
\begin{center}
\includegraphics[width=8cm,angle=0]{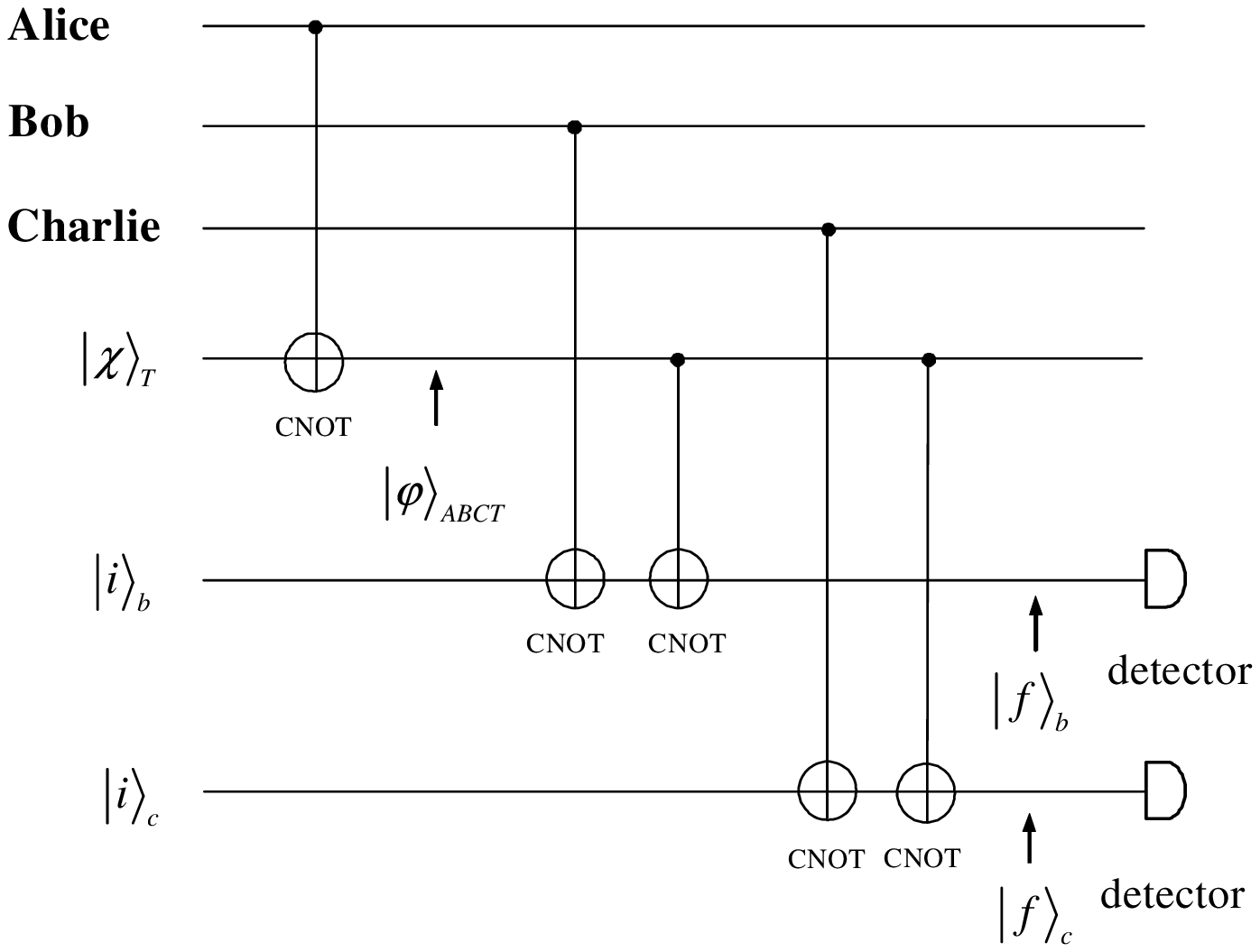} \label{f1}
\caption{ The principle of the multiparty quantum remote secret conference scheme
based on quantum encryption with GHZ states. CNOT: controlled not gate.}
\end{center}
\end{figure}

With quantum memory \cite{storage}, we can modify this MQRSC
protocol to transmit the secret message directly and securely,
without generating the same private classical key and then encrypt
the message, similar to QSDC \cite{two-step,QOTP,Wangc,Wangcoc}. To
this end, the three conferees, Alice, Bob and Charlie, first share a
sequence of three-particle GHZ states $\vert
\psi\rangle_{ABC}=\frac{1}{\sqrt{2}}(\vert 000\rangle + \vert
111\rangle)_{ABC}$, and then use them as a quantum key for
encrypting and decrypting the secret message transmitted. We give
out all the steps for this multiparty quantum remote secret
conference as follows.

(1) The three conferees, Alice, Bob and Charlie, share a sequence of three-particle GHZ-state quantum systems in the same state
$\vert \psi\rangle_{ABC}=\frac{1}{\sqrt{2}}(\vert 000\rangle + \vert 111\rangle)_{ABC}$ securely, i.e.,
ordered $N$ GHZ-state tripartite quantum systems. In detail,
one of the three conferees, say Alice, prepares a sequence of three-particle GHZ states
$\vert \psi\rangle_{ABC}=\frac{1}{\sqrt{2}}(\vert 000\rangle + \vert 111\rangle)_{ABC}$. For each GHZ state, Alice
sends the particle $B_j$ ($j=1, 2, \ldots, N$) to Bob and the particle $C_j$ to Charlie, respectively.
For checking eavesdropping, Alice
chooses randomly some of the GHZ states as their samples, and requires Bob and Charlie to measure their
correlated particles by choosing randomly
the MB $\sigma_x$ or the MB $\sigma_y$. Also Bob and Charlie announce in public their MBs and the outcomes of the measurements.
Here $\vert +y\rangle=\frac{1}{\sqrt{2}}(\vert 0\rangle + i\vert 1\rangle)$ and
$\vert -y\rangle=\frac{1}{\sqrt{2}}(\vert 0\rangle - i\vert 1\rangle)$, i.e.,
\begin{eqnarray}
\vert \psi\rangle_{ABC}&=&\frac{1}{\sqrt{2}}(\vert 000\rangle + \vert 111\rangle)_{ABC}, \nonumber\\
&=& \frac{1}{\sqrt{2}}[(\vert +y\rangle_A\vert -y\rangle_B + \vert
-y\rangle_A\vert +y\rangle_B)\vert +x\rangle_C \nonumber\\
&& + (\vert +y\rangle_A\vert +y\rangle_B + \vert -y\rangle_A\vert
-y\rangle_B)\vert -x\rangle_C.\nonumber\\
\end{eqnarray}
When both Bob and Charlie choose the MB $\sigma_x$ or the MB $\sigma_y$,
Alice chooses the MB $\sigma_x$ to measure her particle $A$ as well, otherwise,
Alice measures her particle with the MB $\sigma_y$. In this way, the three conferees always obtain the correlated
outcomes of their measurements on the samples if there is no eavesdropper,
similar to the two-step protocol \cite{two-step} and BBM92 QKD protocol \cite{BBM92}.
In a noise channel, they can obtain securely a short sequence of GHZ states with entanglement purification.

(2) The three conferees encrypt and decrypt their secret message directly with their quantum key. In detail, a conferee,
say Alice wants to send her secret message to the other two conferees, Bob and Charlie.
She first prepares ordered $N$ traveling particles $S_T: [T_1, T_2, \ldots, T_N]$,
and then encrypts the $j$-th ($j=1,2, \ldots, N$) traveling particle $T_j$
by using a controlled-not (CNOT) gate with the particle $A_j$ in her quantum key, shown in Fig.1.
Suppose the state of a traveling particle $T_j$ is
$\vert \chi \rangle_T= \alpha\vert 0\rangle+ \beta\vert 1\rangle$. Here $\vert \chi \rangle_T \in \{\vert 0\rangle, \vert 1\rangle\}$,
i.e., $\alpha \beta=0$. After the CNOT operation done by Alice with the particle $A_j$ as the control qubit
and the traveling particle $T_j$ as
the target qubit, the state of the quantum system composed of the particles $A_j$, $B_j$, $C_j$ and $T_j$ becomes
\begin{eqnarray}
\vert \varphi\rangle_{ABCT}&=&\frac{1}{\sqrt{2}}(\alpha\vert 0\rangle_A \vert 0\rangle_B \vert 0\rangle_C \vert 0\rangle_T
+ \beta\vert 0\rangle_A \vert 0\rangle_B \vert 0\rangle_C \vert 1\rangle_T\nonumber\\
&+& \alpha\vert 1\rangle_A \vert 1\rangle_B \vert 1\rangle_C \vert 1\rangle_T
+ \beta\vert 1\rangle_A \vert 1\rangle_B \vert 1\rangle_C \vert 0\rangle_T). \nonumber\\
\end{eqnarray}
Alice sends the traveling particles $S_T$ to Bob. For reading out
the information on the traveling particles $S_T$, Bob prepares a
sequence of auxiliary particles $S_b: [b_1, b_2, \ldots, b_N]$ whose
states are initially $\vert i\rangle_b=\vert 0\rangle$, and takes a
CNOT gate on an auxiliary particle $b_j$ and a particle $B_j$ in the
quantum key by using the particle $B_j$ as the control qubit.
Moreover, Bob takes another CNOT gate on the traveling particle
$T_j$ and the auxiliary particle $b_j$ by using the traveling
particle as the control qubit. Thus the state of the auxiliary
particle $b_j$ is changed into the one $\vert f\rangle_b$ as the
same as the original state of the traveling particle $T_j$. Whether
the state $\vert \chi\rangle_T$ is $\vert 0\rangle$ or $\vert
1\rangle$ which represent the bit values 0 and 1 in the secret
message respectively, Bob can read out this information with a
measurement $\sigma_z$ on the auxiliary particle $b_j$.
Simultaneously, Bob sends the traveling particles $S_T$ to a next
conferee, say Charlie.

Certainly, on one hand, Charlie can also do the same operations as those done by
Bob to read out the information about Alice's secret message, shown in Fig.1.
On the other hand, Charlie can measure the traveling particles $S_T$
after he takes a CNOT gate on each traveling particle $T_j$ and the particle $C_j$ in the quantum key by using the particle $C_j$
as the control qubit. As all the quantum systems in the quantum key are in the GHZ state
$\vert \psi\rangle_{ABC}=\frac{1}{\sqrt{2}}(\vert 000\rangle + \vert 111\rangle)_{ABC}$, Charlie can recover the
original state of each traveling particle $T_j$ and then read out the information about Alice's message directly.

(3) The three conferees use their quantum key repeatedly to encrypt and decrypt their secret message in next round
if they confirm that their quantum
communication is secure. For checking the security of their quantum communication, the three conferees
 can choose a subset of quantum systems in their quantum key as samples to analyse their error rate, same as the process
 for sharing a sequence of GHZ states securely in the step 1. Certainly, the new quantum key is shorter a little than the original one.

This MQRSC protocol is secure if the quantum key is secure as it is just a quantum one-time pad
cryto-system \cite{Gisin,QOTP} in this time. None can read out the secret message carried by the traveling
particles $S_T$ if he does not know the information about the quantum key as each traveling particle $T_j$ is
randomly in the states $\vert 0\rangle$ and $\vert 1\rangle$ after the conferee Alice
encrypts it with her particle $A_j$ and a CNOT gate, i.e,  the density matrix for a traveling particle $T$ is $
\rho _{T}=\frac{1}{2}\left(
\begin{array}{cc}
1 & 0  \\
0 & 1
\end{array}
\right)$ whether its original state is $\vert 0\rangle$ or $\vert 1\rangle$. Thus this MQRSC protocol can be made to be secure.

It is straightforward to generalize these two MQSRC schemes to the case with $M$ legitimate conferees, say Alice, Bob, Charlie,
$\ldots$, and Mac. For the first MQSRC scheme, Alice, in this time, prepares an $M$-particle GHZ state
\begin{eqnarray}
\vert \Psi\rangle_{AB\ldots M}=\frac{1}{\sqrt{2}}(\vert 000\ldots 0\rangle + \vert 111 \ldots 1)_{AB\ldots M},
\end{eqnarray}
and sends the particles $B$, $C$, $\dots$, and $M$ to Bob, Charlie,
$\ldots$, and Mac, respectively. Each of the conferees takes the MBs
$\sigma_z$ and $\sigma_x$ to measure his particle with the
probabilities $1-p$ and $p$, respectively. Similar to the case with
three conferees, the probability that all the conferees obtain the
same outcomes with the MB $\sigma_z$ in principle is $(1-p)^M$. Then
the probability that the $M$ remote conferees obtain their raw key,
$p'_{rk}$ without eavesdroppers is
\begin{eqnarray}
p'_{rk}=(1-\sqrt[M]{\frac{r}{2}})^M p'_t p'_d,
\end{eqnarray}
where $p'_t$ and $p'_d$ are the total probabilities of the transmission among the $M$ conferees
and their detectors, respectively. Here
\begin{eqnarray}
p'_t=\prod_{l=2}^{M} p'_{tl},\,\,\,\,\,\,   p'_d =\prod_{l=1}^{M} p'_{dl},
\end{eqnarray}
where $p'_{dl} \leq 1$ is  the probability of the detector of the
$l$-th conferee, $p'_{tl} < 1$ is the probability of the
transmission between the $(l-1)$-th conferee and the $l$-th
conferee. Obviously, the probability $p'_{rk}$ decreases largely
with the increase of the number $M$.

For generalizing our second MQRSC scheme to the case with $M$ conferees, all the conferees should first share securely
a sequence of $M$-particle GHZ states
$\vert \Psi\rangle_{AB\ldots M}=\frac{1}{\sqrt{2}}(\vert 000\ldots 0\rangle + \vert 111 \ldots 1)_{AB\ldots M}$.
This task can be accomplished with the same way as that in the case with three conferees. In the process of encrypting
and decrypting their secret messages, Alice first encrypts her secret message on the traveling particles $S_T:  [T_1,T_2,\ldots T_N]$
with the particles $S_A: [A_1,A_2,\ldots A_N]$ and CNOT gates, and then sends the particles $S_T$ to Bob.
Bob decrypts the secret message
with his auxiliary particles $S_b$ and the particles $S_B: [B_1,B_2,\ldots B_N]$ in the quantum key,
and measures the auxiliary particles
with the MB $\sigma_z$, not the traveling particles $S_T$. He sends the particles $S_T$ to a next conferee,
say Charlie, shown in Fig.1.
All the other conferees just repeat the operations done by Bob except for the last one, Mac. After taking a CNOT
gate on each traveling particle $T_j$ and his particle $M_j$ in the quantum key by using the particle $M_j$
as the control qubit, Mac reads out Alice's  message by measuring the traveling particles $S_T$ with the MB $\sigma_z$.
In order to use their
quantum key repeatedly, all the conferees should check the security of their quantum communication by
sampling some of the quantum systems in their quantum key for analysing their error rate,
similar to that in the case with three conferees.

Compared with quantum secret report \cite{dengreport} and quantum
broadcast communication \cite{wangbroadcast}, each of the legitimate
conferees can send his message to the other conferees, not the case
in which only the agents can send their secret messages to a boss in
a one-way direction \cite{dengreport} or a special one can send his
message to his agents \cite{wangbroadcast}. In our second MQRSC
scheme, only the traveling particles $S_T$ are transmitted among the
$M$ legitimate conferees after they share securely their quantum
key, a short sequence of GHZ states which can be used repeatedly. As
almost all the particles can be used to transmit the secret message,
the intrinsic efficiency for qubits in these two MQRSC schemes
approaches the maximal value. Thus they both are optimal.

In summary, we have presented two multiparty quantum remote secret
conference schemes with GHZ states. One is based on the same
classical key generated with $M$-particle GHZ states and the
measurements with two biased measuring bases. The other is based on
encrypting and decrypting with a quantum key, a sequence of GHZ
states $\vert \Psi\rangle_{AB\ldots M}=\frac{1}{\sqrt{2}}(\vert
000\ldots 0\rangle + \vert 111 \ldots 1)_{AB\ldots M}$. As their
intrinsic efficiency for qubits approaches the maximal value, both
are optimal.

%\end{CJK*}

\end{document}